\documentclass{PoS}

\newcommand\PSb{\Phi_B}

\newcommand\POWHEG{{\tt POWHEG}}
\newcommand\MCatNLO{{\tt MC@NLO}}
\newcommand\MiNLO{{\tt MiNLO}}
\def\sss{\scriptscriptstyle}
\newcommand\as{\alpha_{\sss\rm S}}

\newcommand{\ttbar}{\ensuremath{t \bar t}}
\newcommand{\ttbardecay}{\ensuremath{t \bar t\,\otimes\,}decay}

\newcommand\TTBAR{\ttbar}
\newcommand\TTBARDEC{\ttbardecay}
\newcommand\BBFL{$b \bar b 4\ell$}
\newcommand\BBFLRES{\BBFL{}}
\newcommand\Phirad{\Phi_{\mathrm{rad}}}

\def\nn{\nonumber}

\title{Latest developments in the simulation of final
states involving top-pair and heavy bosons}

\ShortTitle{Latest MC developments for simulating top-pair and VV production}

\author{Emanuele Re\thanks{This research project has been supported by a Maria Sk\l{}odowska-Curie Individual Fellowship of the European Commission's Horizon 2020 Programme under contract number 659147 PrecisionTools4LHC.}~\thanks{Preprint numbers: CERN-TH-2017-095, LAPTH-Conf-012/17}\\~\\
CERN, Theoretical Physics Department, Geneva, Switzerland\\
LAPTh, CNRS, Universit\'e Savoie Mont Blanc, Annecy-le-Vieux, France\\
E-mail: \email{emanuele.re@lapth.cnrs.fr}}

\abstract{I give an overview of recent progress in the simulation of
  final states involving top-quarks and vector bosons pair. First I'll
  discuss the recently found solutions needed to simulate fully
  differential top pair production ($pp\to b\bar{b}+4$ leptons) at
  NLO+PS accuracy, retaining off-shellness and interference effects
  exactly. In the second part, I'll review the \MiNLO{}
  (Multi-scale Improved NLO) method, and then show a recent
  application, namely the simultaneous NLO+PS description of $W^+W^-$
  and $W^+W^-+1$ jet production.}

\FullConference{38th International Conference on High Energy Physics\\
		3-10 August 2016\\
		Chicago, USA}

\begin{document}

In this article I will review a couple of recent results on the
inclusion of higher-order QCD corrections to the Monte Carlo
simulation of final states involving top-pair and heavy electroweak
bosons. In section~\ref{sec:tt} I will focus on the recent progress
achieved in the matching of QCD NLO corrections with parton shower
simulations (NLO+PS) for the process $pp\to W^+ W^- b\bar{b}$, whereas
in section~\ref{sec:ww} I will discuss the NLO+PS merging of $W^+W^-$
and $W^+W^-$+1 jet production using the \MiNLO{}
method.\footnote{Unless otherwise stated, throughout this document, we
  indicate with ``$W$'' the lepton-neutrino final-state pair arising
  from a $W$ bosons, \emph{i.e.} $W$ bosons are treated as unstable,
  have a finite width and they decay leptonically.}

\section{top-pair production}
\label{sec:tt}

It is known that having an accurate simulation of the process $pp\to
W^+W^-b\bar{b}$ is important for several reasons at the LHC, for
instance to measure the top quark mass, or to have an unified
treatment of $t\bar{t}$ and the so-called ``single-top $Wt$''
production.

At fixed order in QCD, $W^+W^-b\bar{b}$ hadronic production is very
well known and much studied.
Despite the fully differential cross section has been known for
several years at
NLO~\cite{Denner:2012yc,Bevilacqua:2010qb,Heinrich:2013qaa,Frederix:2013gra,Cascioli:2013wga},
with the exception of a first study appeared
in~\cite{Garzelli:2014dka}, Monte Carlo NLO+PS event generators
addressing all the issues related to the complete simulation of this
final state started to be available only recently in the \POWHEG{}
approach~\cite{Campbell:2014kua,Jezo:2015aia,Jezo:2016ujg}.
In the rest of this section I'll focus on these issues (and solutions
thereof) within the \POWHEG{} approach, although substantial work in this
direction is also pursued within the \MCatNLO{} matching scheme, and
complete results for single-top t-channel production in the 4-flavour
scheme were published in ref.~\cite{Frederix:2016rdc}.

The problem with the simulation of $W^+W^-b\bar{b}$ production and the
inclusion of finite width effects can be stated as follows: unless
special care is taken,
the intermediate top-quark virtuality is not preserved among different
parts of the computation, leading to the evaluation of matrix elements
at phase space points which have different top virtualities. When this
happens, three problems will in general occur:
\begin{enumerate}
  \item at the level of computing NLO corrections with a
    subtraction method, the cancellation of collinear singularities
    associated to gluon emission off final-state $b$-quarks can become
    delicate, eventually failing when approaching the narrow-width
    limit.
  \item when the hardest radiation is generated in \POWHEG{},
    the phase-space region associated to final-state gluon emission
    off the $b$-quark is handled by a mapping that, in general, does
    not preserve the virtuality of the intermediate resonance.
    Unless $m^2_{bg} \ll \Gamma_t E_{bg}$,
    real and Born matrix elements ($R$ and $B$, respectively) will not be on the resonance peak at the same time,
    hence the ratio $R/B$ in the \POWHEG{} Sudakov
    can become large when $R$ is on peak and $B$ is
    not, yielding a spurious ``Sudakov suppression''.
  \item further problems can arise during the parton-showering
    stage: from the second emission onward, the shower should be
    instructed to preserve the mass of the resonances. This could be
    done easily if there was an unique mechanism to ``assign'' the
    radiation to a given resonance. For processes where
    interference(s) is(are) present, no obvious choice is possible.
\end{enumerate}

An intermediate solution to the previous issues was presented in
ref.~\cite{Campbell:2014kua}, where a fully consistent NLO+PS
simulation for $W^+W^-b\bar{b}$ production was obtained in the
narrow-width limit, and off-shellenss and interference effects were
implemented in an approximate way. I refer to the original paper, or
to the review~\cite{Re:2016psv}, for more details. Here it suffices to say
that, by using the narrow-with approximation to compute NLO
corrections, production and decay can be clearly
separated (no interference arises), thereby allowing a
non-ambiguous ``resonance assignment'' for final-state radiation, as
well as the use of an improved (``resonance aware'') subtraction
method, where radiation in the decay is generated by first boosting
momenta in the resonance rest-frame. In this way, $B$ and $R$ are
always evaluated with the same virtuality for the intermediate
resonance,
so that the subtraction can be safely performed, and no spurious
Sudakov suppression can arise.

More recently, a general solution to include off-shellness and
interference effects in the\\{\ttfamily \POWHEG{}} approach was
proposed in ref.~\cite{Jezo:2015aia}, and later applied to the
$W^+W^-b\bar{b}$ process in ref.~\cite{Jezo:2016ujg}, where
matrix elements were obtained using {\ttfamily
  OpenLoops}~\cite{Cascioli:2011va}.  The main new concept introduced
in~\cite{Jezo:2015aia} is that one separates all contributions to the
cross section into terms with definite resonance structure,
\emph{i.e.} each term should only have peaks associated to a given
resonance structure (``resonance history''). For $W^+W^-b\bar{b}$
production, for instance, one has two types of resonance histories:
one where at least one $s$-channel top-propagator appears (this
includes both doubly- and single-resonant contributions) and another
associated to non-resonant production. By means of projectors
$\Pi_{f_b}$ built by combining Breit-Wigner like functions $P^{f_b}$,
a partition of the unit can be constructed, so that a given Born (and
virtual) partonic subprocess $B$ can be separated into contributions $B_{f_b}$
that are, individually, dominated by one and only one resonance history
(labeled by $f_b$):\footnote{For simplicity we suppressed the labels
  $F_b$ and $F_r$ used in~\cite{Jezo:2015aia}, which represent the
  ``bare'' structure, \emph{i.e.} the flavour of external partons of
  Born and real matrix elements. Moreover in~\cite{Jezo:2015aia} the
  symbols $f_b$ and $f_r$ represent a ``full'' structure, since they
  label a given resonance history related to a given set of external
  partons, \emph{i.e.} they also contain the explicit information on
  the external partons, which we are suppressing in this document.}
\begin{equation}
\label{eq:tt_born}
B(\PSb) = \sum_{f_b} B_{f_b}(\PSb) \equiv \sum_{f_b} \Pi_{f_b}(\PSb) B(\PSb),\ \ \mbox{where }\ \ \Pi_{f_b}(\PSb) = \frac{P^{f_b}(\PSb)}{\sum_{f'_b}P^{f'_b}(\PSb)}\,.
\end{equation}
Since each Born matrix element is separated according to resonance
histories, one needs to set-up a similar mechanism for real matrix
elements, such that, eventually, each projected real matrix element
can be associated to an unique resonance history, with a counterpart
in the corresponding list of Born's ones. As usual, real matrix
elements also need be separated according to their collinear
singularities: to this end, one requires that a collinear region
$\alpha_r$ is admitted only if the two collinear partons \emph{both}
arise either from the same resonance, or from the hard
interaction. This separation is achieved schematically as
\begin{equation}
\label{eq:tt_real}
R = \sum_{\alpha_r} R_{\alpha_r},\ \ \mbox{where }\ \ R_{\alpha_r} = \frac{P^{f_r} d^{-1}(\alpha_r)}{\sum_{f'_r}(P^{f'_r} \sum_{\alpha'_r}d^{-1}(\alpha'_r))} R\,.
\end{equation}
In eq.~(\ref{eq:tt_real}), $f_r$ denotes a given resonance history
assignment for $R$, $d(\alpha_r)\to 0$ when the collinear region
$\alpha_r$ is approached, and the sums in the denominator run only on
the possible resonance histories $f'_r$ present in $R$, and on the
compatible singular regions $\alpha'_r$ associated to a given $f'_r$: hence 
a given $R_{\alpha_r}$ becomes dominant only if the collinear partons
of region $\alpha_r$ have the smallest $k_t$ \emph{and} the
corresponding resonance history $f_r$ is the closest to its mass shell.

The above prescriptions allowed to build a \POWHEG{} generator able to
simulate processes with intermediate resonances, keeping all
finite-width effects and interferences. In fact, having separated each
contribution as explained above, for the singular regions associated
to a radiation in a resonance decay, it becomes now possible to safely
use the ``resonance-aware'' subtraction method developed
in~\cite{Campbell:2014kua}, thereby avoiding the mismatches mentioned
at the beginning of this section. Similarly, because an index
$\alpha_r$ is naturally associated to the hardest radiation generated
by \POWHEG{}, it's possible to unambiguously assign the radiation to a
given resonance, preventing the parton shower to distort the mass of
the resonances.\footnote{I want to mention that another technical but
  crucial issue was addressed in~\cite{Jezo:2015aia}, related to the
  computation of soft-collinear contributions to be added to to the
  virtual terms. In~\cite{Campbell:2014kua}, these terms were computed
  independently for production and for each radiating resonance decay,
  and in different frames. This posed no problem, because in the
  narrow-width limit no interferences are present. When interferences
  are present, this is clearly no longer possible, and a substantial
  generalization of the subtraction scheme adopted in {\ttfamily
    POWHEG} was worked out in~\cite{Jezo:2015aia}, leading to the
  development of a new framework, dubbed {\ttfamily POWHEG-BOX-RES}.}

\begin{figure}
  \includegraphics[width=0.49\textwidth]{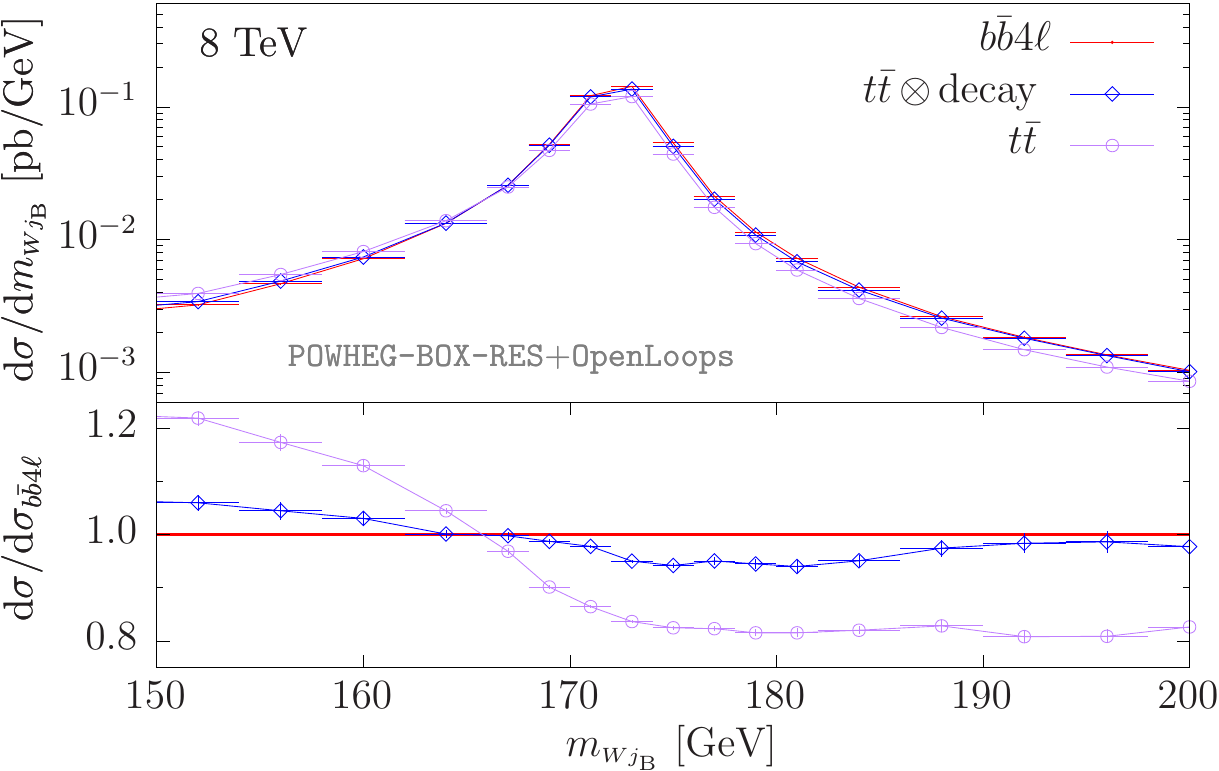}~
  \includegraphics[width=0.49\textwidth]{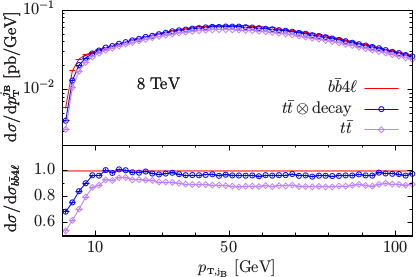}
  \caption{Invariant mass of the $W$ and hardest $b$-jet (left panel)
    and transverse momentum of the hardest $b$-jet (right panel), at
    the $\sqrt{S}=8$ TeV LHC. Figures taken from
    ref.~\cite{Jezo:2016ujg}.}
  \label{fig:tt}
\end{figure}

The left panel of Fig.~\ref{fig:tt}
shows the differences in shape of the reconstructed top peak obtained
using three tools, namely the new generator of
ref.~\cite{Jezo:2016ujg} (\BBFLRES{}), the resonance-improved
generator based on an approximate treatment of off-shell
effects~\cite{Campbell:2014kua} (\TTBARDEC{}), and the original
generator~\cite{Frixione:2007nw} based on on-shell NLO matrix elements
for $t\bar{t}$ production (\TTBAR{}). As expected, the ``\BBFLRES{}''
and ``\TTBARDEC{}'' generators are fairly consistent, especially close
to the resonance peak, whereas the ``\TTBAR{}'' generator shows larger
deviations. The $p_T$ spectrum of the hardest $b$-jet is instead shown
in the right panel of fig.~\ref{fig:tt}, without imposing any
particular cuts. The shape difference at small $p_T$ can be attributed
to the fact that the ``Wt'' contribution is missing (approximate) in
the ``\TTBAR{}'' (``\TTBARDEC{}'') generator, whereas it's fully taken
into account in the ``\BBFLRES{}'' one.

\section{$W$-boson pair production}
\label{sec:ww}
The study of vector boson pair-production is central for the LHC
Physics program. Not only is $W^+W^-$ production measured to access
anomalous gauge couplings, but it's also an important background for
several searches, notably for those where the $H\to W^+W^-$ decay is
present. For these and other similar reasons, it is important to have
flexible and fully realistic theoretical predictions that allow to
simultaneously model, with high accuracy, the production of $W^+W^-$,
inclusively as well as in presence of jets. The methods aiming at this
task are usually referred to as ``NLO+PS merging''. NLO+PS merging for
$pp\to VV$+jet(s) was achieved using the {\ttfamily
  MEPS@NLO}~\cite{Hoeche:2012yf,Cascioli:2013gfa} and {\ttfamily
  FxFx}~\cite{Frederix:2012ps,Alwall:2014hca} methods. In this
section, I'll review the \MiNLO{} formalism and show how it was used
to merge at NLO the processes $pp\to W^+W^-$ and $pp\to W^+W^-+$
jet~\cite{Hamilton:2016bfu}.

The \MiNLO{} (Multi-scale Improved NLO)
procedure~\cite{Hamilton:2012np} was originally introduced as a
prescription to a-priori choose the renormalization ($\mu_R$) and
factorization ($\mu_F$) scales in multileg NLO computations: since
these computations can probe kinematic regimes involving several
different scales, the choice of $\mu_R$ and $\mu_F$ is indeed
ambiguous, and the \MiNLO{} method addresses this issue by
consistently including CKKW-like
corrections~\cite{Catani:2001cc,Lonnblad:2001iq} into a standard NLO
computation. In practice this is achieved by associating a
``most-probable'' branching history to each kinematic configuration,
through which it becomes possible to evaluate the couplings at the
branching scales, as well as to include (\MiNLO{}) Sudakov form
factors (FF). This prescription regularizes the NLO computation also
in the regions where jets become unresolved, hence the \MiNLO{}
procedure can be used within the \POWHEG{} formalism to regulate the
$\bar{B}$ function for processes involving jets at LO.

In a single equation, for a $q\bar{q}$-induced process as $W^+W^-$
production, the \MiNLO{}-improved \POWHEG{} $\bar{B}$ function reads:
\begin{equation}
  \label{eq:ww-minlo}
  \bar{B}_{\,\tt WWJ-MiNLO} =  \as(q_T) \Delta^2_q(q_T,M_X)   
  \Big{[}
  B ( 1-2\Delta^{(1)}_q(q_T,M_X) ) + 
  \as V(\bar{\mu}_R) +\as \int d\Phirad R
  \Big{]}\nn\,,
\end{equation}
where $X$ is the color-singlet system ($WW$ in this case), $q_T$ is
its transverse momentum, $\bar{\mu}_R$ is set to $q_T$, and
$\Delta_q(q_T,Q)=\exp\Big{\{}-\int_{q^2_T}^{Q^2}\frac{dq^2}{q^2}\frac{\as(q^2)}{2\pi}
\Big{[} A_{q}\log\frac{Q^2}{q^2} + B_{q} \Big{]}\Big{\}}$ is the
\MiNLO{} Sudakov FF associated to the jet present at LO. Convolutions
with PDFs are understood, $B$ is the leading-order matrix element for
the process $pp\to X+1$ jet (stripped off of the strong coupling), and
$\Delta_{q}^{(1)}(q_T,Q)$ (the $\mathcal{O} (\as)$ expansion of
$\Delta_q$) is removed to avoid double counting. We also notice that
$\bar{B}_{\,\tt WWJ-MiNLO}$ is a function of $\Phi_{X+j}$,
\emph{i.e.} the phase space to produce the $X$ system and an extra
parton, which can be arbitrarily soft and/or collinear.

In ref.~\cite{Hamilton:2012rf} it was also realized that, if $X$ is a
color singlet, upon integration over the full phase space for the
leading jet, one can formally recover NLO+PS accuracy for the process
$pp\to X$ by properly applying \MiNLO{} to NLO+PS simulations for
processes of the type $pp\to X+1$ jet.\footnote{The idea has been
  generalized recently in ref.~\cite{Frederix:2015fyz}.} Besides
setting $\mu_F$ and $\mu_R$ equal to $q_T$ in all their occurrences,
the key point is to include at least part of the
Next-to-Next-to-Leading Logarithmic (NNLL) corrections into the
\MiNLO{} Sudakov form factor, namely the $B_2$ term: by omitting it,
the full integral of eq.~(\ref{eq:ww-minlo}) over $\Phi_{X+j}$, albeit
finite, differs from $\sigma_{pp\to X}^{NLO}$ by a relative amount
$\as(M_X)^{3/2}$, thereby hampering a claim of NLO accuracy.

The $B_2$ coefficient is process-dependent, and formally also a
function of $\Phi_{X}$, because part of it stems from the 1-loop
correction to the $pp\to X$ process. For Higgs, Drell-Yan, and $VH$
production, these 1-loop corrections can be expressed as form factors:
$B_2$ becomes just a number as its dependence upon $\Phi_{X}$ disappears, and
the analogous of eq.~(\ref{eq:ww-minlo}) can be easily
implemented~\cite{Hamilton:2012rf,Luisoni:2013kna}. For diboson
production, the situation is more delicate. First, extracting $B_2$
for the $WW$ case is more subtle, as the virtual corrections to the
$pp\to WW$ process don't factorize on the Born squared amplitude,
hence $B_2=B_2(\Phi_{X})$. As a consequence, a mismatch between
different phase spaces becomes apparent, because in
eq.~(\ref{eq:ww-minlo}) $\bar{B}_{\,\tt WWJ-MiNLO}$ depends upon
$\Phi_{X+j}$, whereas $B_2$ needs to be computed as a function of
$\Phi_{X}$. In ref.~\cite{Hamilton:2016bfu} these two issues were
handled as follows:
\begin{itemize}
\item to compute $B_2$ we started from the relatively simple
  expression used for the Drell-Yan case, and replaced its
  process-dependent part $[V/B]^{{\tt DY}}=C_F(\pi^2-8)$ with the
  corresponding term for $WW$ production:
  $[V/B]^{WW}(\Phi_{WW})=V^{WW}(\Phi_{WW})/B^{WW}(\Phi_{WW})$.
\item in order to evaluate $B_2$, we defined on an event-by-event
  basis a projection of the $WW+1$ jet state onto a $WW$ one, using
  the FKS mapping relevant for initial-state radiation as implemented
  in the {\ttfamily POWHEG BOX}~\cite{Frixione:2007vw}. For real
  emission events, a similar mapping was used. In all cases, in the
  $q_T\to 0$ limit, the effect of these projections on the final state
  kinematics smoothly vanishes, making sure that the precise numerical
  determination of $B_2$ is affected only beyond the required
  accuracy.
\end{itemize}

In ref.~\cite{Hamilton:2016bfu} we have built a \POWHEG{} generator
for the $pp\to W^+W^-+1$ jet process, and upgraded it with \MiNLO{},
according to the aforementioned procedure.  We worked in the 4-flavour
scheme, including exactly the vector bosons' decay products, as well
as finite-width effects and single-resonant contributions. Tree-level
matrix elements were obtained with an interface to\\{\tt
  MadGraph\,4}~\cite{Alwall:2007st,Campbell:2012am}, whereas one loop
corrections were computed with {\tt
  GoSam\,2.0}~\cite{Cullen:2014yla,Luisoni:2013kna}.

\begin{figure}
  \includegraphics[width=0.49\textwidth]{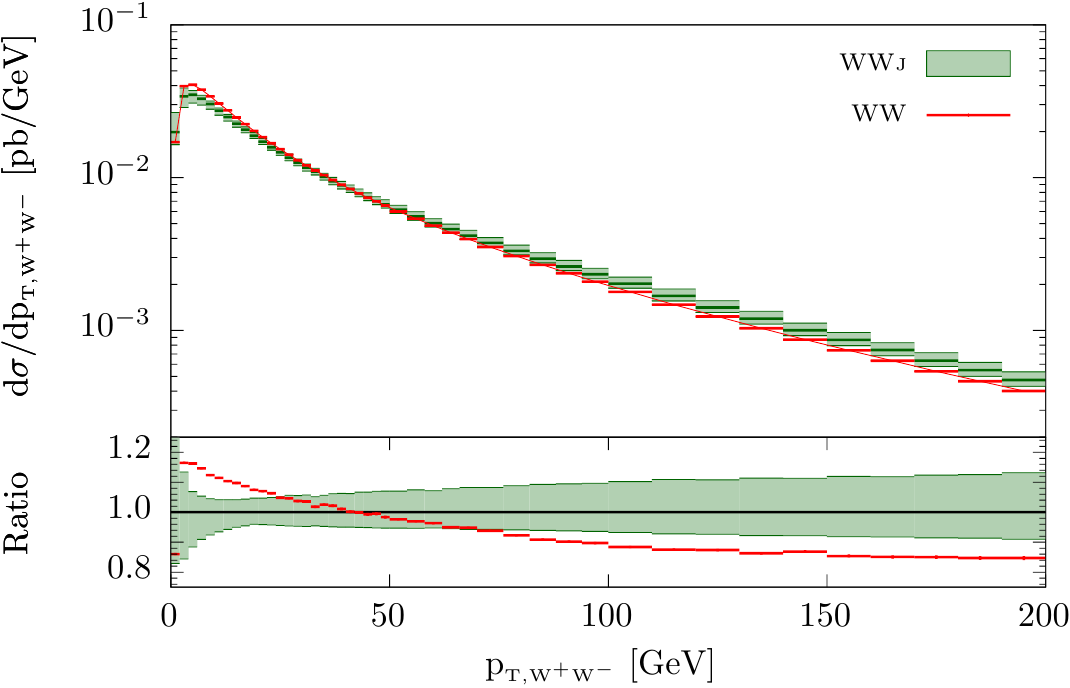}~
  \includegraphics[width=0.49\textwidth]{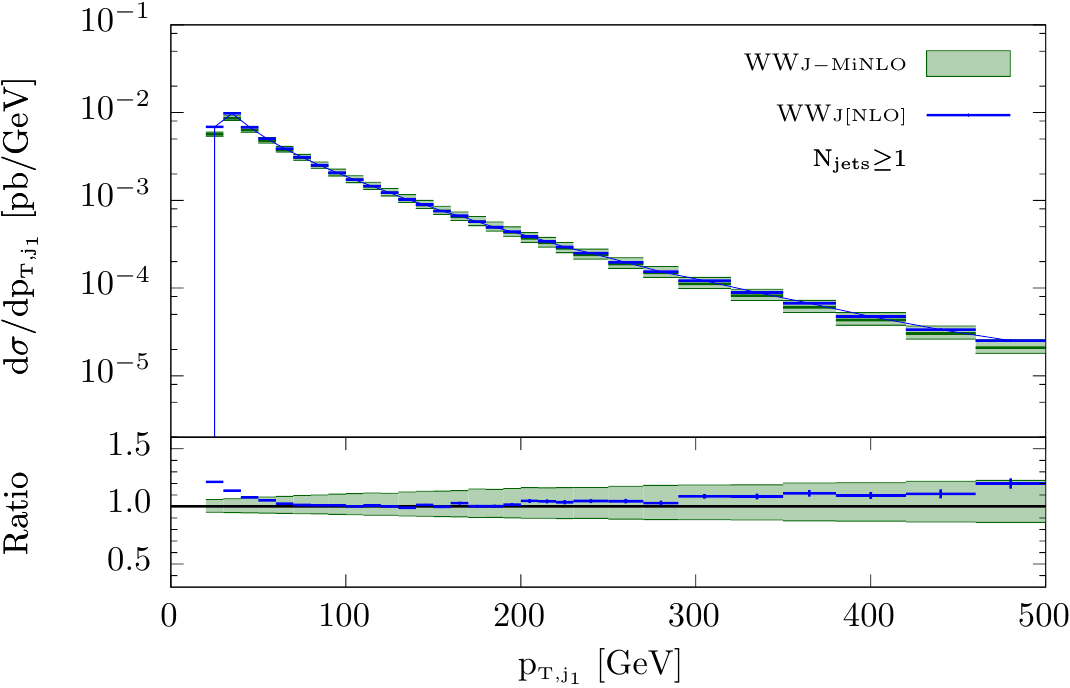}
  \caption{Transverse momenta of the $W^+W^-$ system (left panel) and
    of the leading jet (right panel), at the $\sqrt{S}=13$ TeV
    LHC. The green curve is the \MiNLO{} prediction (central value
    with uncertainty) from ref.~\cite{Hamilton:2016bfu}, the red one
    is obtained using the original \POWHEG{}
    generator~\cite{Melia:2011tj}, and the blue line is a partonic NLO
    result. Figures taken from ref.~\cite{Hamilton:2016bfu}.}
  \label{fig:ww}
\end{figure}

The left panel of Fig.~\ref{fig:ww}
shows the transverse momentum spectrum of the $WW$ system as obtained
with the {\tt WWJ-MiNLO} generator against the one obtained with the
original \POWHEG{} generator for $pp\to
W^+W^-$~\cite{Melia:2011tj}. The importance of NLO corrections is
manifest in the high-$p_T$ tail, whereas the differences at small
$p_T$ can be attributed to the differences among the {\ttfamily
  POWHEG} and the {\ttfamily MiNLO} Sudakovs.  The right panel shows
instead a comparison for the leading jet $p_T$ spectrum between the
parton-level computation $pp\to WW+1$ jet at NLO, and the \MiNLO{}
result. This observable is formally described with the same
accuracy (NLO) by both predictions, as shown in the plot. The effect
of resumming collinear logarithms at small $p_T$ is reflected in the
difference between \MiNLO{} and the pure NLO, where no resummation is
included. At high $p_{T,j}$, the small differences are due to the fact
that different central values for the $\mu_R$ and $\mu_F$ scales are
used, namely $\mu=p_{T,WW}$ for \MiNLO{} (as by prescription) and
$\mu=m_{WW}$ at NLO.

It will be interesting to improve further the {\ttfamily WWJ-MiNLO}
generator by including the effect of $gg$-induced contributions (an
NLO+PS study for the $ZZ$ case was performed in
ref.~\cite{Alioli:2016xab}) and, ultimately, matching it to the
differential NNLO computation of ref.~\cite{Grazzini:2016ctr}.

\end{document}